\documentclass[prd,twocolumn,preprintnumbers,superscriptaddress,nofootinbib]{revtex4}
\pdfoutput=1

\usepackage[english]{babel}
\usepackage{slashed}
\usepackage{amsmath,amsfonts}
\usepackage[]{units}
\usepackage{multirow}
\usepackage{color}
\usepackage{enumitem}
\usepackage{hyperref}
\usepackage[all]{hypcap}
\usepackage{array}
\usepackage{float}
\usepackage{graphicx}
\usepackage[caption=false]{subfig}
\usepackage{slashed}

\usepackage[dvipsnames]{xcolor}

\newcommand{\tr}[1]{\text{tr}(#1)}
\newcommand{\DM}{\text{FP}}
\newcommand{\mPl}{\text{m}_\text{Pl}}

\allowdisplaybreaks
\begin{document}

\preprint{ULB-TH/17-16}

\title{Self-interacting spin-2 dark matter}

\author{Xiaoyong Chu}
\email{Xiaoyong.Chu@oeaw.ac.at }
\affiliation{Institute of High Energy Physics, Austrian Academy of Sciences, Nikolsdorfer Gasse 18, 1050 Vienna, Austria
}

\author{Camilo Garcia-Cely}
\email{Camilo.Alfredo.Garcia.Cely@ulb.ac.be}
\affiliation{Service de Physique Th\'eorique - Universit\'e Libre de Bruxelles, Boulevard du Triomphe, CP225, 1050 Brussels, Belgium}

\hypersetup{
 pdftitle={Self-interacting Spin-2 Dark Matter},
 pdfauthor={Xiaoyong Chu, Camilo Garcia-Cely}
}

\begin{abstract}
Recent developments in bigravity allow one to construct consistent theories of interacting spin-2 particles that are free of ghosts. In this framework, we propose an elementary spin-2 dark matter candidate with a mass well below the TeV scale. We show that, in a certain regime where the interactions induced by the spin-2 fields do not lead to large departures from the predictions of general relativity, such a light dark matter particle typically self-interacts and undergoes self-annihilations via 3-to-2 processes. We discuss its production mechanisms and also identify the regions of the parameter space where self-interactions can alleviate the discrepancies at small scales between the predictions of the collisionless dark matter paradigm and cosmological N-body simulations.
\end{abstract}

\maketitle

\section{Introduction}
\label{sec:intro}

\begin{figure}[t]
\includegraphics[height=0.28\textwidth]{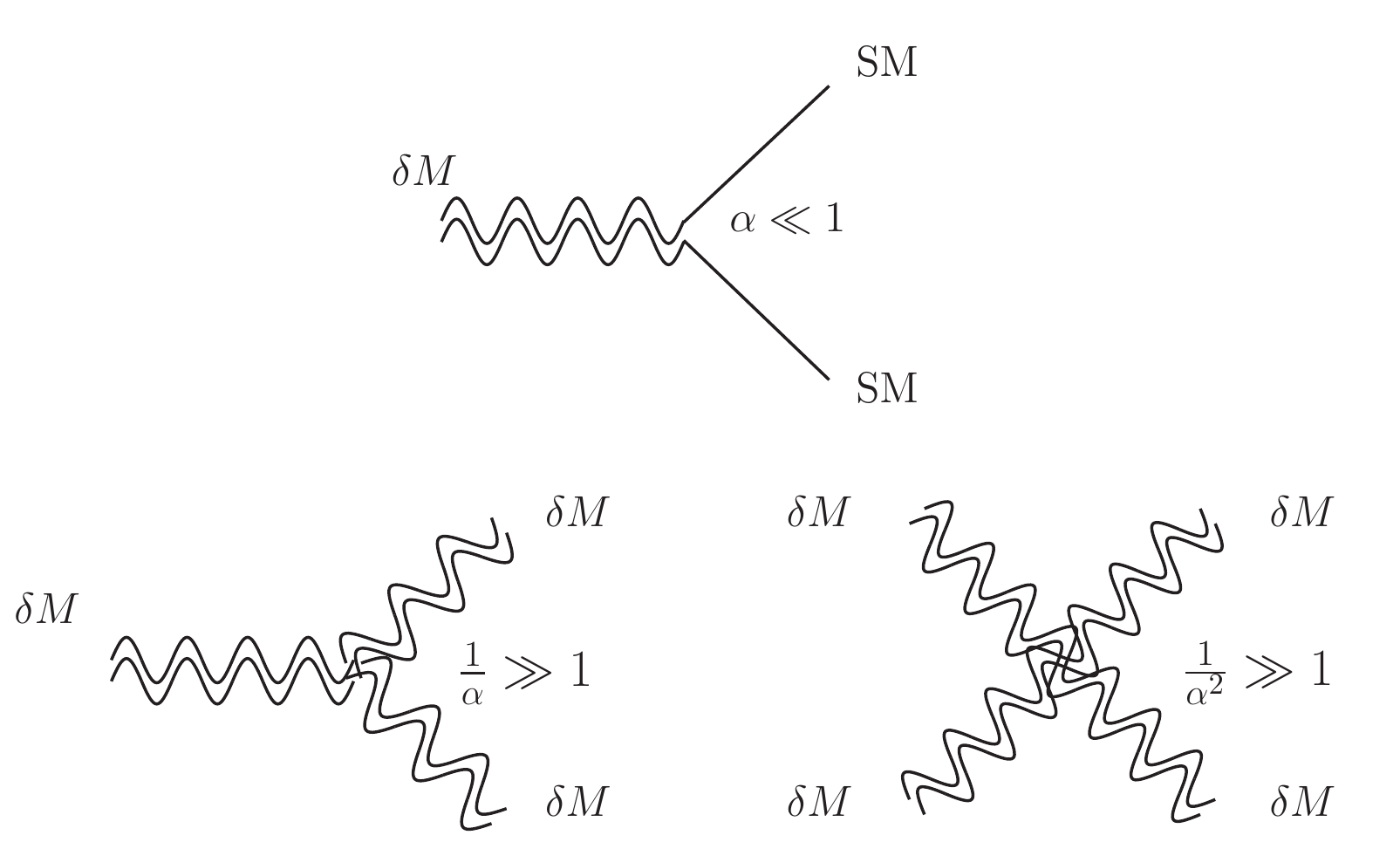}
\caption{
The portal communicating the spin-2 DM particle $\delta M$ to the SM fields is proportional to $\alpha$, whereas its self-interaction vertices are inversely proportional to powers of $\alpha$. 
}
\label{fig:portal}
\end{figure}

Multiple astrophysical and cosmological observations provide strong evidence for the existence of dark matter (DM). Nevertheless, as of today, we remain agnostic of its particle nature. In particular, we do not know its spin. Several theories containing DM particles with spin 0, 1/2 and 1 have been constructed as it is well known how to properly describe them by following the success of the Standard Model (SM) of particle physics~\cite{Olive:2016xmw}. This is in sharp contrast with the case of spin-2 particles. In fact, until very recently it was not possible to consistently study the interactions of such particles due to the presence of Boulware-Deser
ghosts in all known theories of interacting massive spin-2 particles~\cite{Boulware:1973my}. Bimetric theories of gravity, also known as bigravity, 
 have recently overcome this difficulty~\cite{Hassan:2011hr,Hassan:2011tf,Hassan:2012wr}. For a review of these developments, see Ref.~\cite{Schmidt-May:2015vnx}. All this motivates us to further explore the possibility that DM is an elementary particle of spin-2 in the context of bigravity. Such scenarios have been first considered in Refs.~\cite{Babichev:2016hir, Babichev:2016bxi, Aoki:2016zgp} (see also~\cite{Maeda:2013bha}).

A crucial observation that we would like to highlight in this article is that, in such bimetric theories, the spin-2 DM particle naturally self-interacts. 
Concretely, the strength of the interactions among massive spin-2 particles is inversely proportional to powers of their portal to the SM sector, which is typically small in order to recover the predictions of general relativity (GR), as illustrated in Fig.~\ref{fig:portal}. As we will see, the fact that spin-2 DM self-interacts has at least two interesting phenomenological consequences. 

On the one hand, because of the smallness of the portal to the SM sector, spin-2 DM cannot reach kinetic equilibrium with the SM bath in the early Universe. In fact, it cannot efficiently annihilate or decay into SM particles either. Accordingly, DM cannot be produced via the standard freeze-out mechanism. Nevertheless, it can be produced via freeze-in 
when the self-interactions are small~\cite{McDonald:2001vt,Hall:2009bx}, either at the end of inflation or during preheating/reheating periods, as originally pointed out in Refs.~\cite{Babichev:2016hir, Babichev:2016bxi, Aoki:2016zgp}. In each case, Cosmic Microwave Background observations put upper bounds on the isocurvature perturbations, the gravitational-wave energy spectrum, and the reheating (RH) temperature $T_\text{RH}$, strongly disfavoring the low-mass region of spin-2 DM. 
For instance, if DM is nonthermally produced from the feeble portal interaction to SM particles during reheating,
the bound on $T_\text{RH}$ leads to the DM masses roughly above 1 TeV~\cite{Babichev:2016bxi,Babichev:2016hir}. In contrast, as proposed in this work, one more production mechanism naturally arises for light spin-2 DM when its self-interactions are sufficiently large.
The latter, if strong enough, lead to a thermalization of the dark sector after it is produced from annihilations of SM particles. 
Then, the relic density is set by the freeze-out of 3-to-2 processes in a dark bath colder than the SM particles. 
Such a possibility of generating the DM abundance via number-changing processes have been widely studied in the literature~\cite{Dolgov:1980uu,Carlson:1992fn,Hochberg:2014dra,Yamanaka:2014pva,Hochberg:2014kqa,Bernal:2015bla,Bernal:2015lbl,Lee:2015gsa,Choi:2015bya,Hansen:2015yaa,Bernal:2015xba, Kuflik:2015isi,Hochberg:2015vrg,Choi:2016hid,Pappadopulo:2016pkp,Farina:2016llk,Choi:2016tkj,Dey:2016qgf,Cline:2017tka, Choi:2017mkk, Choi:2017zww}. All this opens up the possibility of having thermal spin-2 DM below the GeV scale.\footnote{By analogy with the axion, ``the misalignment mechanism''~\cite{Preskill:1982cy,Abbott:1982af,Dine:1982ah} has also been invoked \cite{Marzola:2017lbt} very recently in order to generate the observed abundance of light spin-2 DM. It applies to DM lighter than ${\mathcal O}$(0.1)\,eV, being different from the keV--GeV mass range of interest for self-interacting DM.}

On the other hand, for some regions of the parameter space, the self-interaction cross section in DM halos can be as large as $\unit[1]{cm^2/g}$. In fact, 
while there is no conclusive evidence suggesting that DM particles scatter off each other,
N-body simulations of collisionless DM fail to properly describe small-scale DM-dominated objects such as the dark (sub)halos that host dwarf or low-surface-brightness galaxies~\cite{Walker:2011zu,deNaray:2011hy,BoylanKolchin:2011de, BoylanKolchin:2011dk}. 
Interestingly, postulating the existence of DM self-scatterings in these objects provides a plausible solution to these problems~\cite{Spergel:1999mh}. For a recent review of these small-scale anomalies as well as of models of self-interacting DM, see Ref.~\cite{Tulin:2017ara}. In view of this, we will also investigate the parameter space where the strength of the spin-2 DM self-interactions is enough to alleviate such small-scale problems. 

This article is organized as follows: in Sec.~\ref{sec:DM} we provide a brief introduction to the ghost-free bimetric theory, when the massive spin-2 particle it predicts plays the role of DM. In Sec.~\ref{sec:self} we discuss DM self-interactions. 
Production mechanisms for our self-interacting DM candidate are presented in Sec.~\ref{sec:production}. Finally, we conclude in Sec.~\ref{sec:conclusion}. Appendix~\ref{sec:app} is devoted to the details of the self-interaction cross section for massive spin-2 particles.

\section{Spin-2 Dark Matter}
\label{sec:DM}

Any spin-2 particle with no other quantum numbers generally mixes with the graviton of GR. Since the latter is described as perturbations of the space-time metric, it is natural to assume that a theory with two metrics, $f_{\mu\nu}$ and $g_{\mu\nu}$, will describe two spin-2 states. In fact, this is the case of bimetric theories, which are described by the action\footnote{In this work, we closely follow the notations of Ref.~\cite{Babichev:2016bxi}.} 
\begin{eqnarray}
{\cal S}
 &=& m_g^2 \int d^4x \left(\sqrt{-g} R(g) + \alpha^2 \sqrt{-f} R(f) \right.\nonumber\\ &&\left. -2 \alpha^2 m_g^2 \sqrt{-g}\, V(S; \beta_n)\right)+{\cal S}_\text{matter}\,. \label{eq:action}
\end{eqnarray}
The first and second terms in this equation are the Einstein-Hilbert (EH) actions for the metrics $f$ and $g$, respectively. Notice that the $\alpha$ parameter is a ratio between the couplings associated with these tensor fields. The third term then describes the interactions between both metrics. This part of the action is determined~\cite{deRham:2010kj, Hassan:2011zd} by requiring no Boulware-Deser ghosts~\cite{Boulware:1973my}. There, $\beta_n$ stands for five coupling constants and $S$ is the square root of $g^{-1}f$, that is $S^\rho{}_\sigma S^\sigma{}_\nu = g^{\rho\mu}f_{\mu\nu}$. The potential reads
\begin{eqnarray}
V(S;\beta_n) &=& \beta_0+\beta_1 \,S^{\mu_1}{}_{\mu_1} + \beta_2 \,S^{\mu_1}{}_{[\mu_1} S^{\mu_2}{}_{\mu_2]} \,\nonumber\\ &&+ \beta_3 \,S^{\mu_1}{}_{[\mu_1} S^{\mu_2}{}_{\mu_2} S^{\mu_3}{}_{\mu_3]}+\beta_4 \,\text{det}{S}\,. 
\end{eqnarray}
Such a potential exhibits the symmetry $\sqrt{-g} V(S; \beta_n) = \sqrt{-f} V(S^{-1}; \beta_{4-n})$ and therefore the gravitational action is symmetric under the exchange of the two metrics together with $\alpha\leftrightarrow \alpha^{-1}$ and $\beta_n\leftrightarrow \beta_{4-n}$. Nevertheless, this is not the case for the matter action, which may only minimally couple the matter fields to one of the metrics in order not to introduce ghosts (see \cite{Schmidt-May:2015vnx} and references therein). Without loss of generality, we thus have
\begin{eqnarray}
{\cal S}_\text{matter}= \int d^4x \sqrt{-g} \, {\cal L}_\text{matter} (g,\Phi)\,,
\end{eqnarray} 
where $\Phi$ is a generic field of the Standard Model. 

\begin{figure}[t]
\includegraphics[height=0.17\textwidth]{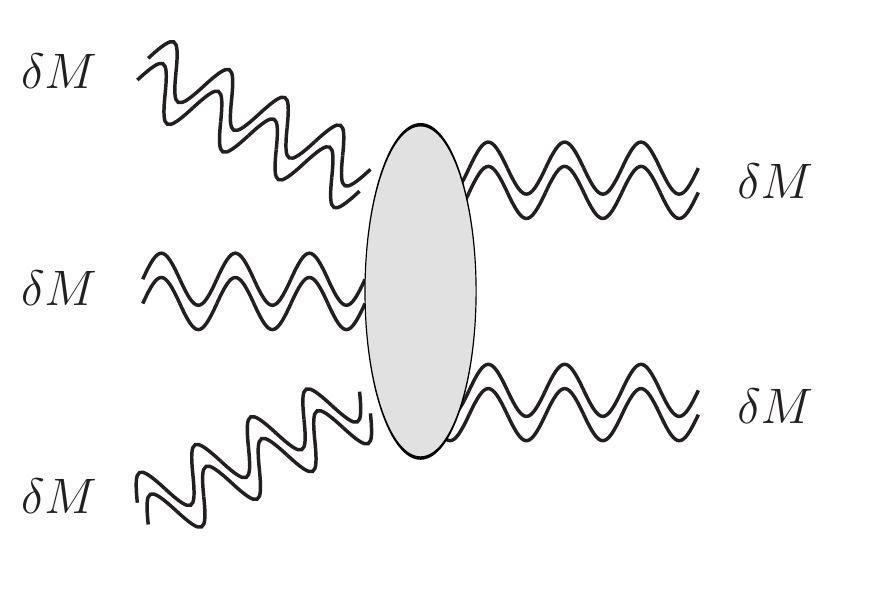}
\vspace{-0.5cm}
\caption{
3-to-2 annihilation process of spin-2 DM particles. 
}
\label{fig:3to2}
\end{figure}

Spin-2 particles can be considered as linear fluctuations of $f$ and $g$ around the same background metric $\overline{g}$.\footnote{In general, the background solutions only need to be proportional to each other, but the proportionality constant can be absorbed by redefining the $\alpha$ and the $\beta_n$ parameters, as adopted here.} Concretely, the decomposition is given by~\cite{Hassan:2012wr}
\begin{align}
g_{\mu\nu} \equiv \overline{g}_{\mu\nu}+h_{\mu\nu} \,,&& f_{\mu\nu} \equiv \overline{g}_{\mu\nu}+\ell_{\mu\nu} \,,
\end{align}
where the fields $h$ and $l$ mix with each other. From Eq.~\eqref{eq:action} it follows that the corresponding spin-2 mass eigenstates, $\delta G$ and $\delta M$, can be obtained via the transformation
\begin{eqnarray}
\label{eq:h}
h_{\mu\nu} &=& \frac{1}{m_g \sqrt{1+\alpha^2} }\left(\delta G_{\mu\nu}-\alpha \delta M_{\mu\nu}\right)\,, \\
\label{eq:l}
l_{\mu\nu} &=& \frac{1}{m_g \sqrt{1+\alpha^2} }\left(\delta G_{\mu\nu}+\alpha
^{-1} \delta M_{\mu\nu}\right)\,, 
\end{eqnarray} 
where the parameter $\alpha$ determines the mixing between the linear fluctuations $h$ and $l$. Note that each of the two tensors has one mass dimension. The linear and quadratic pieces of the action~\eqref{eq:action} in terms of these fields are 
\begin{eqnarray}
\label{eq:actionlinear}
{\cal S}&\supset& \int d^4x \sqrt{-\overline{g}}\left({\cal L}_\Lambda+{\cal L}_2^\text{EH}(\delta G) + {\cal L}_2^\text{EH}(\delta M)\right.\nonumber\\ && \left. -\frac{m_\DM^2}{4} \left(\tr{\delta M^2}-\tr{\delta M}^2\right)\right.\,\nonumber\\&& \left.-\frac{1}{\mPl} \left(\delta G_{\mu\nu}-\alpha \delta M_{\mu\nu}\right) T^{\mu\nu}\right)\,,
\end{eqnarray} 
where the indices are raised and lowered with the background metric, $\mPl$ is the Fierz-Pauli mass associated with the field $\delta$M, and $\mPl$ is the reduced Planck mass. Here, ${\cal L}_\Lambda$ includes terms involving the cosmological constant, which is given by
\begin{eqnarray}
\Lambda &=& \alpha^2 m_g^2 (\beta_0+3\beta_1 +3\beta_2 +\beta_3)\,\\
&=& m_g^2 (\beta_4+3\beta_3 +3\beta_2 +\beta_1)\,.
\end{eqnarray}
Also, ${\cal L}^\text{EH}_2$ is the kinetic term for spin-2 particles as derived from the EH action (see Eq.~\eqref{eq:kinHE}). In addition, there are
\begin{eqnarray}
\mPl&\equiv&m_g\sqrt{1+\alpha^2}\,,\label{eq:planckmass}\\
m_\DM &\equiv& \mPl \sqrt{\beta_1+2\beta_2+\beta_3}\,,\\
T_{\mu\nu} &\equiv& -\frac{1}{\sqrt{-g}} \frac{\partial (\sqrt{-g}{\cal L}_\text{matter})}{\partial g^{\mu\nu}} \label{eq:energymomentum}\,.
\end{eqnarray}
Lagrangian terms for $\delta M$ beyond the quadratic level of Eq.~\eqref{eq:actionlinear} are presented in Appendix~\ref{sec:app}, which also lists the vertices of Fig.~\ref{fig:portal} and those giving rise to the process of Fig.~\ref{fig:3to2}. 

Many comments are already in order at this stage:
\begin{itemize}
\item The action in Eq.~\eqref{eq:actionlinear} describes two spin-2 particles. One of them, $\delta G$, is massless and is thus the graviton. The other particle, $\delta M$, is described by a Fierz-Pauli Lagrangian~\cite{Pauli:1939xp} and is therefore a spin-2 particle with mass $m_\DM$. 
\item Both the cosmological constant and the mass of the spin-2 particle are determined by the coupling constants $\beta_n$. As noted in Refs.~\cite{Babichev:2016bxi,Babichev:2016hir}, a hierarchy between both scales necessarily requires one to tune such constants. We will not address this issue in this work and simply assume that the cosmological constant takes the observed value and that the spin-2 particle mass lies above $\mathcal O$(1) MeV. As a result, for calculations concerning DM particles, it is a good approximation 
to assume that the background metric is flat.
\item Both fields couple to the energy-momentum tensor associated with the metric $g$, as given by Eq.~\eqref{eq:energymomentum}. Moreover, the coupling of the graviton to $T_{\mu\nu}$ defines the reduced Planck mass in Eq.~\eqref{eq:planckmass}. With this, the coupling associated with the massive spin-2 particle is $\alpha/\mPl$. 
Thus, the interaction between the dark and the SM sectors is mediated by the gravitational portal times the coupling constant $\alpha$ (see Fig.~\ref{fig:portal}). 
\end{itemize}

Since this theory modifies conventional GR, the laws of gravity generally change. One must therefore ensure that the corresponding predictions are in agreement with those of GR at the scales where the latter has been tested. The simplest way to achieve that is by assuming $\alpha \ll 1$~\cite{Hassan:2012wr}. In this way, the gravitational interactions mediated by the tensor $\delta M$ are much weaker than those associated with the graviton. For more detailed discussions, we refer the reader to~\cite{Babichev:2016bxi}.

Interestingly, for $\alpha \ll 1$, the massive spin-2 particle cannot decay or annihilate into SM particles at an appreciable rate, as Fig.~\ref{fig:portal} suggests. 
In fact, it has been shown~\cite{Babichev:2016bxi} that its lifetime is given by
\begin{equation}
\tau_{\delta M } \equiv \frac{1}{\Gamma(\delta M \to \text{SM}\text{SM})} \sim \frac{80 \pi\mPl^2}{\alpha^2 m_\DM^3}\,,
\label{eq:ind} 
\end{equation}
which can be much longer than the age of the Universe. 
Furthermore, as shown in Refs.~\cite{Babichev:2016bxi,Babichev:2016hir}, Eq.~\eqref{eq:action} implies that the vertex $\delta G \delta G \delta M$ is zero. From this, it follows that the $\delta M$ boson cannot decay into two gravitons either. Hence, for feeble values of $\alpha$, the massive spin-2 is stable on cosmological scales, and thus can serve as an excellent DM candidate.

\section{Self-Interactions}
\label{sec:self}

\begin{figure*}[t]
\includegraphics[width=0.50\textwidth]{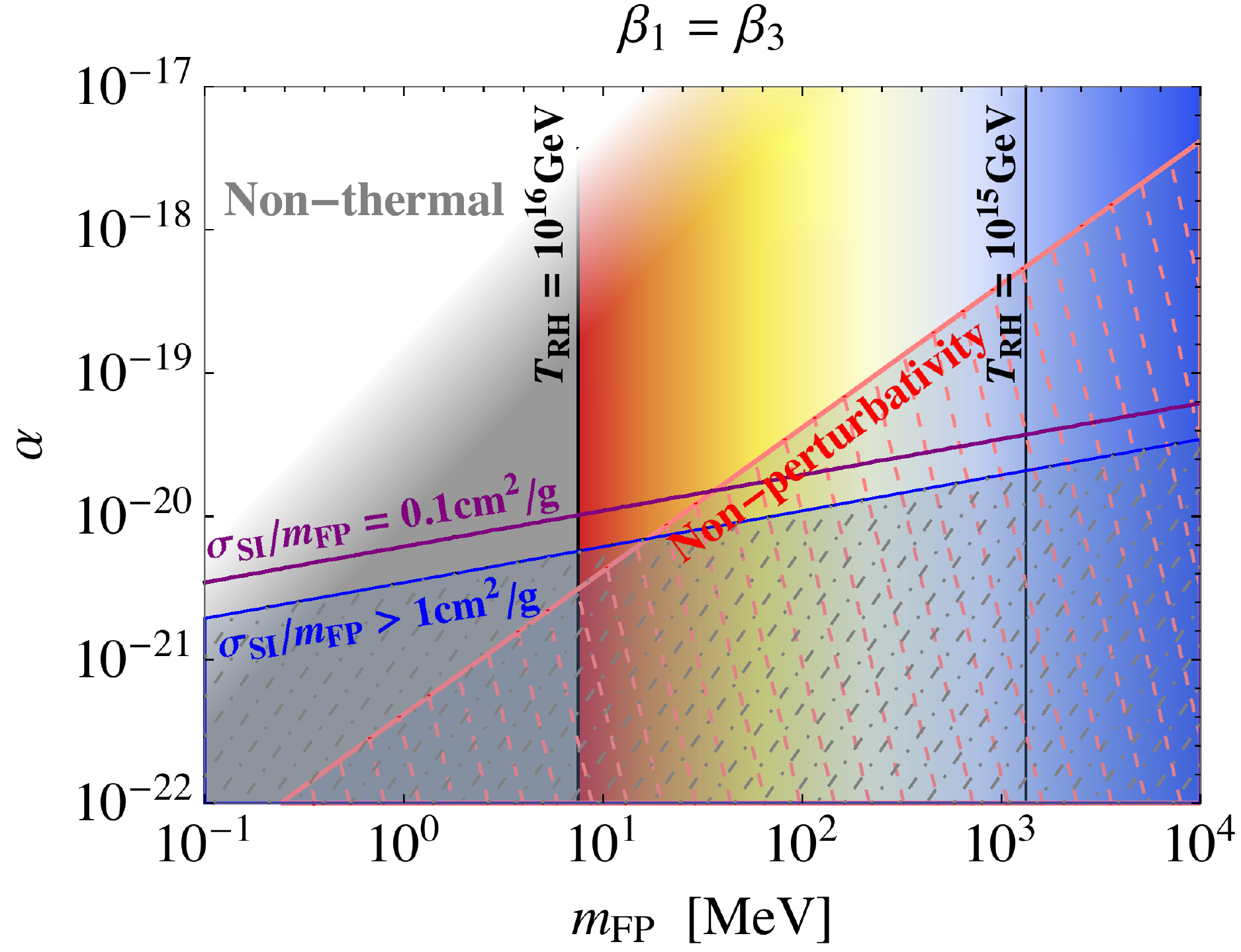}~~~
\includegraphics[width=0.49\textwidth]{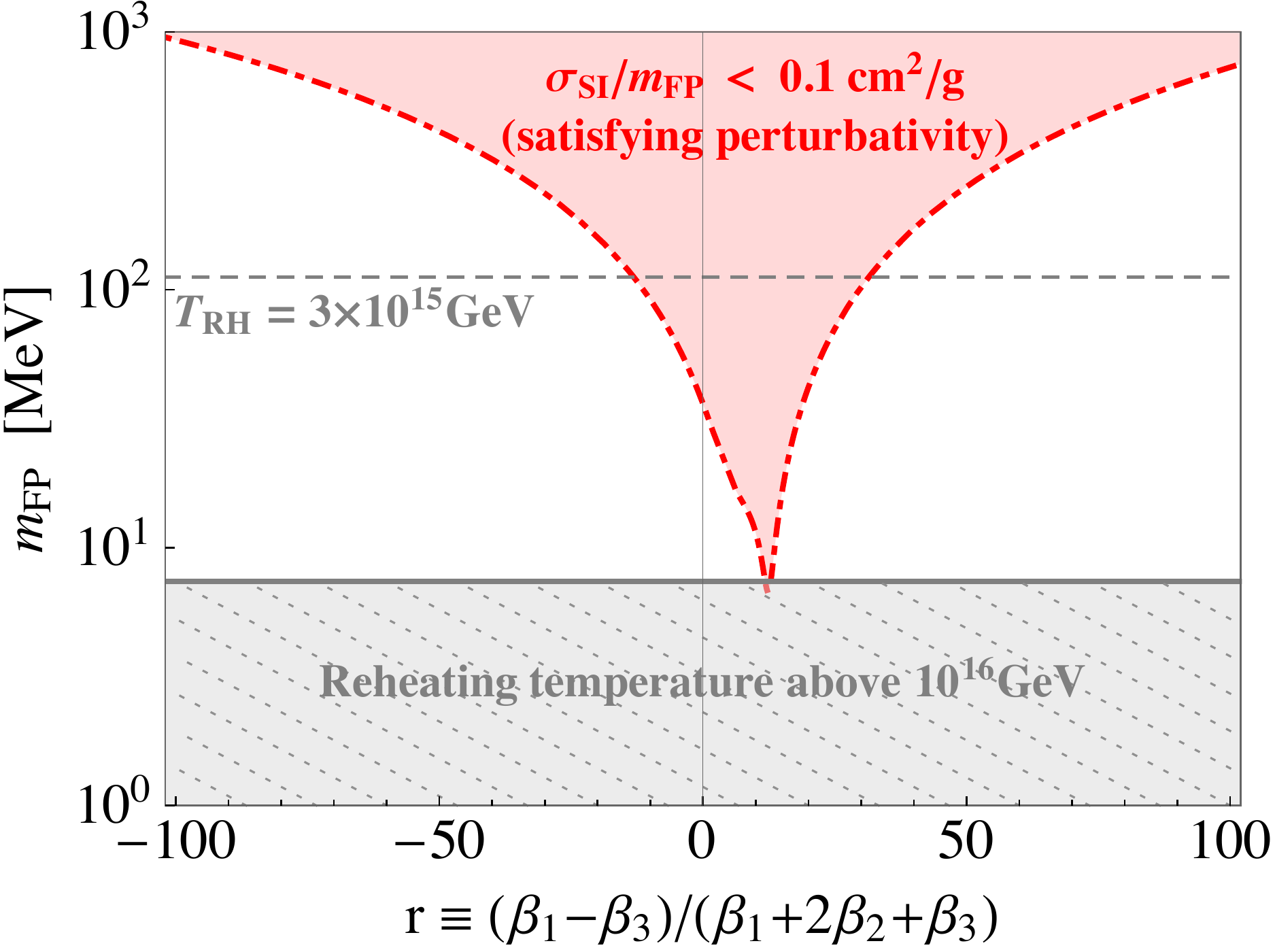}
\caption{
\emph{Left}: parameter space for self-interacting spin-2 DM when $\beta_1= \beta_3$. The color gradient corresponds to the required reheating temperature $T_\text{RH}$ to produce the observed relic abundance. The left gray region, corresponding to $T_\text{RH} \ge 10^{16}$\,GeV is considered to be strongly disfavored. The nonperturbativity region is given by $\alpha \mPl \ge m_\text{FP}$. \emph{Right}:  white region shows the DM mass range of interest as a function of $r$ to address small-scale problems (i.e., $\sigma_\text{SI}/m_\text{FP} \ge 0.1$\,cm$^2$/g) while being allowed by perturbativity requirement (e.g., the $r=0$ vertical line corresponds to the purple line in left panel). All hatched regions have been excluded in both panels. 
}
\label{fig:plot1}
\end{figure*}

The gravitational part of the action in Eq.~\eqref{eq:action} treats both metrics $g$ and $f$ (or equivalently, $h$ and $l$) symmetrically. Moreover, 
in the limit $\alpha\ll1$, Eqs.~\eqref{eq:h} and \eqref{eq:l} give
\begin{align}
\label{eq:hllimit}
h_{\mu\nu} &= \frac{\delta G_{\mu\nu}}{\mPl} +\mathcal{O}(\alpha^1)\,,&& 
l_{\mu\nu} = \frac{1}{\alpha}\frac{\delta M_{\mu\nu}}{\mPl}+\mathcal{O}(\alpha^0)\,. 
\end{align} 
 Therefore, loosely speaking, we can say that the self-interactions of the DM field, $\delta M$, are the same as those of graviton, $\delta G$, but enhanced by powers of $1/\alpha$. On the other hand, the interactions of the $\delta M$ boson with the SM fields are suppressed by $\alpha$ in comparison with the interactions of the graviton $\delta G$. We can conclude that bimetric theories provide a natural framework for self-interacting DM of spin-2 in the decoupling regime where gravitational laws do not significantly deviate from those of GR.

In order to be more precise, we present the vertices involving three and four $\delta M$ bosons to the leading order in $\alpha$ in Appendix~\ref{sec:app}. We find that the cubic and the quartic vertices are proportional to $\alpha^{-1}$ and $\alpha^{-2}$, respectively.
 With them, we calculate the two-body self-interaction (SI) cross section for nonrelativistic DM as determined by the vertices shown in Fig.~\ref{fig:portal},
\begin{eqnarray}
\sigma_\text{SI} &=& \frac{ (\beta_1+2\beta_2+\beta_3)^2}{ \pi 
 \alpha^4 m_\DM^2} F\left(\frac{\beta_1-\beta_3}{\beta_1+2\beta_2+\beta_3}\right)\,,
\end{eqnarray}
where $F$ is a dimensionless polynomial expression, explicitly given in Eq.~\eqref{eq:Fr}. We would like to remark that vertices involving $\delta G$ do not contribute to the cross section at the leading order in $\alpha$.

The appearance of the small parameter $\alpha$ in the denominator legitimately raises the question of the validity of the perturbative expansion for calculating scattering rates. This issue has been systematically studied in Ref.~\cite{Babichev:2016bxi}. They found that, for processes with typical energy $E$, the perturbativity condition is given by $E \lesssim\alpha \,\mPl$. In the case of self-interactions of nonrelativistic DM, the typical energy is just the DM mass. In our case, such a condition is then 
\begin{equation}
\label{eq:pertur}
\alpha^2 \gtrsim m^2_\DM/\mPl^2 = \beta_1+2\beta_2+\beta_3 \,. 
\end{equation}

Regarding the current experimental limits on $\sigma_\text{SI}$, the nonobservation of an offset between the mass distributions of DM and stars in the Bullet Cluster sets an upper bound on the self-interacting cross section, $\sigma_\text{SI} /m_\text{FP}<1.25$\,cm$^2$/g at $68\%$~C.L.~\cite{Randall:2007ph}. Moreover, recent N-body simulations suggest that this value must not be much larger than $0.1$\,cm$^2$/g at cluster scales to fit the measured density profile of cluster Abell 2667~\cite{Elbert:2016dbb}. On the other hand, it has been also pointed out that the angular dependence of the cross section might affect the observable consequences associated with the DM self-scattering~\cite{Kummer:2017bhr}. For the purpose of this work, we will not further discuss this, and simply take a practical constraint $\sigma_\text{SI} /m_\text{FP}<1$\,cm$^2$/g below, in our numerical calculations. The region where this constraint is satisfied is shown in the left panel of Fig.~\ref{fig:plot1}, in the plane of $\alpha$ vs. $m_\DM$, for benchmark parameters such that $\beta_1=\beta_3$. In the same panel, we also show the perturbativity condition (above the red line) as given by Eq.~\eqref{eq:pertur}. 

As mentioned in the Introduction, self-interacting DM can alleviate some of the tensions between observations and simulations that persist in the collisionless DM paradigm at small scales. As recently summarized in \cite{Tulin:2017ara}, cosmological N-body simulations suggest that a plausible solution is found if there exists DM self-interaction at the order of 
\begin{equation}
\label{eq:sigmaoverm}
\frac{\sigma_\text{SI} }{m_\DM} \sim 0.1-\unit[1]{cm^2/g}\,.
\end{equation}
The region where this condition is fulfilled is  depicted in left panel of Fig.~\ref{fig:plot1} in between the (nearly horizontal) purple and blue lines for $\beta_1=\beta_3$. Notice that the perturbativity bound, being fixed by $\alpha$ and $m_\DM$, is the same for a different choice of $\beta_1$ and $\beta_3$,  while the lines corresponding to constant $\sigma_\text{SI} /m_\text{FP}$ move up or down according to the value $r=(\beta_1-\beta_3)/(\beta_1+2\beta_2+\beta_3)$. This is illustrated by the dashed-dotted red line in the right panel of Fig.~\ref{fig:plot1}, which gives the intersection between the line associated with the perturbativity bound and that for which  $\sigma_\text{SI} /m_\text{FP} =0.1$\,cm$^2$/g.

We conclude that, for $\beta_1=\beta_3$, the DM mass must be below roughly 30~MeV if the small-scale problems are addressed by our spin-2 DM candidate. For general choices of $\beta_i$, the  upper bound on $m_\DM$ can be read from the right panel of Fig.~\ref{fig:plot1}. Unsurprisingly, the $m_\text{FP}$ value of most interest is about tens of MeV, which is usually the case for self-interacting DM without light mediators. 

We would like to remark that, in the present model, the self-interaction cross section for nonrelativistic DM is independent of the DM velocity. 
Reference  \cite{Kaplinghat:2015aga} has found that observations prefer a mildly velocity-dependent cross section if the hypothesis of self-interacting DM correctly accounts for the small-scale problems discussed above. If such a velocity dependence is confirmed by further investigations, the present model would be strongly disfavored as a solution to the small-scale problems.

\section{Production Mechanism}
\label{sec:production}

Throughout this work, we assume that DM is dominantly produced by the transfer of energy from the SM bath to the DM sector during reheating. After the energy transfer becomes suppressed, both sectors completely decouple. How the dark sector evolves thereafter depends on the strength of DM self-interactions, which are discussed in the following paragraphs.\footnote{See Ref.~\cite{Bernal:2015ova} for a comprehensive study of various production mechanisms of self-interacting dark matter candidates.}
\vspace{-0.2cm}
\subparagraph{Freeze-in.} This mechanism was first considered for spin-2 DM with negligible self-interactions in Ref.~\cite{Babichev:2016bxi}.
The Boltzmann equation describing the produced DM abundance, i.e. the ratio of DM number density, $n_\text{DM}$, to the total entropy density $s$, is 
\begin{equation}
	{dY\over dz} = { s\over H z } Y^2_\text{eq} \langle \sigma v \rangle_\text{\tiny DM$\leftrightarrow$SM}\,,\label{Boltz:numberFI}
\end{equation}
where $H$ is the Hubble rate, $Y \equiv n_\text{DM}/ s$ and $z\equiv {m_\DM/T} $. Also, $Y_\text{eq} \equiv n_\text{eq}(m_\DM, T)/s $ represents the thermal abundances of DM as a function of the photon temperature $T$. 
The abundance quickly reaches a constant value $Y_f$, and the relic density is then given by 
\begin{eqnarray}
\Omega_\text{DM} &=&\frac{ Y_f \, s \, m_\DM}{\eta_B n_\gamma m_p}\times \Omega_B\nonumber\\ &\simeq& {0.12\over h^2} \left( \,{Y_f\over 4.38\times 10^{-10}}\right)\left( \,{m_\DM\over \text{GeV} }\right)\,,
\label{eq:OmegaFI}
\end{eqnarray}
where $m_p$ is the proton mass, while $\Omega_B$ and $\eta_B$ are the baryon mass parameter and baryon-to-photon ratio, respectively. Cosmological parameters measured by full-mission Planck data~\cite{Ade:2015xua} have been adopted to obtain the last equality.

For spin-2 DM, the energy transfer mainly takes place via annihilations of SM particles into DM pairs.\footnote{The inverse-decay channel $SM SM \to \delta M$ is negligible due to phase-space suppression for parameter regions considered here.} The corresponding cross section scales as $\langle \sigma v \rangle_\text{\tiny DM$\leftrightarrow$SM}\sim T^2/\mPl^4$~\cite{Babichev:2016bxi}. Notice that the $\alpha$ dependence cancels in the cross section. For instance, the s-channel exchange of $\delta M$ in the annihilation processes have one vertex proportional to $\alpha/\mPl$ and the other proportional to $1/(\alpha \mPl)$. The annihilations are thus more efficient at early times, so that the production process is determined by the reheating temperature, $T_\text{RH}$. Integrating the lhs of Eq.~\eqref{Boltz:numberFI} over $z$ gives 
\begin{equation}
Y_f \simeq \left({36\sqrt{2} \, \xi(3)^2\over \pi^{11/2} \sqrt{g_*}}\right) \left({ T_\text{RH} \over m_\text{Pl}}\right)^{3}\simeq 0.013 \left({ T_\text{RH} \over m_\text{Pl}}\right)^{3}\,,
\label{estimate:nonthermal}
\end{equation}
where $g_*$ is the total effective degrees of freedom at reheating, and the Riemann zeta function $\xi(3)\simeq 1.202$. 
By plugging this and the observed DM relic density into Eq.~\eqref{eq:OmegaFI}, one obtains $
m_\DM \sim \unit[1]{TeV} \left({\unit[10^{15}]{GeV} /T_\text{RH} }\right)^{3}. 
$
As Fig.~\ref{fig:plot1} suggests, unless we choose unphysically large values for the $\beta$ parameters, such DM masses lead to $\sigma_\text{SI}/m_\DM \ll \unit[0.1]{cm^2/g}$. This means that DM self-interactions are irrelevant, as already implicitly assumed in the freeze-in regime~\cite{Bernal:2017kxu}.

\subparagraph{Dark freeze-out.} As mentioned in previous sections, self-interactions can be important for light DM and must be taken into account for calculating the relic density. In fact, after being produced from annihilations of SM particles, DM self-thermalization would naturally take place for sufficiently light DM. 
Furthermore, as a consequence of the chemical equilibration processes shown in Fig.~\ref{fig:3to2}, the total DM number density increases at the price of reducing the kinetic energy per particle compared with the freeze-in regime above.

Hence, here we consider a scenario where a thermal bath of spin-2 DM particles--characterized by the DM temperature $T'$--is produced after the energy-transfer from the SM sector stops and where the freeze-out of the 3-to-2 annihilation processes leads to the observed DM abundance. Describing such a physical system requires two Boltzmann equations. One of them determines the energy transfer by tracking the energy ratio between two sectors, $\rho'/\rho$. This reads
\begin{equation}
	{d(\rho'/\rho)\over d z } = { s^2 \over H z \rho } \, Y^2_\text{eq} \langle \sigma v \cdot \Delta E \rangle_\text{\tiny DM$\leftrightarrow$SM} \,, \label{Boltz:energy}
\end{equation}
where the pair-creation cross section, weighted by energy transferred per process $\Delta E $, can be estimated by $ T^2/\mPl^4 \times 2T$. Correspondingly, after the effective energy-transfer process decouples, the density ratio roughly approaches a constant, namely ${\rho'_\text{de}/ \rho_\text{de}}$. Then this ratio remains the same (up to decoupling of heavy SM particles) while the DM particles are relativistic, even after they thermalize. 
Integration of Eq.~\eqref{Boltz:energy} gives ${\rho'_\text{de}/ \rho_\text{de}}$, which can be used to calculate the temperature of the DM sector via 
\begin{equation}
 {T' \over T} = \left({\rho'_\text{de}/g'_*\over \rho_\text{de} /g_*}\right)^{1\over4} \simeq \left({24\sqrt{2} \,\xi(3)^2 \sqrt{g_*} \over \pi^{11/2} g'_*}\right) ^{1\over4} \left({ T_\text{RH} \over m_\text{Pl}}\right)^{3\over 4}\,, \label{eq:initialratioT}
\end{equation}
where the dark relativistic degrees of freedom $g'_* =5$ for massive spin-2 particles. From now on, we will denote the prefactor in the last term as $\eta^{1 / 4}$ to simplify the expressions below. Numerically, $\eta \simeq 0.187$ if there are no other new particles. Note that the value of $T'/T$ is always very small, so the extra radiation bound from observable primordial abundances is automatically satisfied. 

The second Boltzmann equation describes the evolution of the DM number density
\begin{equation}
	{dY\over dz} = { s\over H z }\left\{Y^2_\text{eq} \langle \sigma v \rangle_\text{\tiny DM$\leftrightarrow$SM} - s Y^2 ( Y- Y'_\text{eq} )\langle \sigma v^2 \rangle_{3 \to 2}\right\} \,,\label{Boltz:number}
\end{equation}
in which $Y_\text{eq}$ is the same as above, and $Y'_\text{eq}= n_\text{eq}(m_\DM, T')/s $, giving the abundance of DM chemical equilibrium corresponding to $T'$. In addition, $\sigma v^2$ stands for the cross section associated with the 3-to-2 process, which chemically equilibrates the dark plasma. As explained in Appendix~\ref{sec:app}, the EH action on the metric $f$ gives rise to quintic vertices scaling as $m_\DM^2/(\alpha^3 \,\mPl^3)$. We thus expect that, for nonrelativistic DM, $\langle \sigma v^2 \rangle \propto m_\DM/(\alpha^6 \mPl^6) = (\beta_1+2\beta_2+\beta_3)^3/(\alpha^6 m_\DM^5) $.

Before discussing the solution of the Boltzmann equations, let us first specify the condition for the dark thermalization induced by the number-changing process of Fig.~\ref{fig:3to2}. This process needs to happen before the DM particles become nonrelativistic. This fact allows one to establish in which part of the parameter space the DM abundance can arise from freeze-out of the 3-to-2 process. 
As a matter of fact, the comparison of the rate for this process and the Hubble rate excludes the white region of Fig.~\ref{fig:3to2} at the top-left corner. There, the 3-to-2 and its inverse process can hardly thermalize the dark sector or produce the observed relic abundance of DM. This means that an extra (nonthermal) production mechanism must be invoked for the parameter space indicated by the white region.

We can now 
estimate the solution of the Boltzmann equation \eqref{Boltz:number} from the conservation of the density or the entropy ratio between the two sectors depending on how DM particles freeze out: 
\begin{itemize}

\item \emph{Relativistic freeze-out: } in this case, the energy density ratio will not change with Universe expansion until freeze-out.
According to Eq.~\eqref{eq:initialratioT}, we expect the freeze-out DM abundance to be approximately given by
\begin{eqnarray}
	Y_f &=& \left({45\xi(3) \over 2\pi^4}\right)\left({T' \over T}\right)^{3} 
\simeq 
 \left({45\xi(3)\eta^{3\over 4} \over 2\pi^4}\right) \left({ T_\text{RH} \over m_\text{Pl}}\right)^{9\over 4} \notag \\
 &\simeq & 0.08\left({ T_\text{RH} \over m_\text{Pl}}\right)^{9\over 4} \,.\label{estimate:relaFO}
\end{eqnarray}

By plugging this result in Eq.~\eqref{eq:OmegaFI}, the observed relic density determines the reheating temperature for a given DM mass. 

\item \emph{Nonrelativistic freeze-out: } here, the energy ratio is not conserved as the DM density undergoes a Boltzmann suppression before the freeze-out. Nevertheless, because the DM particles self-interact, after the energy transfer between both sector stops, 
their entropy ratio is conserved~\cite{Carlson:1992fn}. When the DM particles are relativistic, their entropy ratio is given by $s'/s \sim (\rho'_\text{de}/\rho_\text{de})^{3/4}$. When they become nonrelativistic, the dark entropy determines the DM number density via $s' \simeq m_\DM n_\text{DM}/T'$. Thus, at freeze-out (FO) we expect 
\begin{eqnarray}
Y_f & \equiv &	{n_\text{DM} \over s } \simeq {T'_\text{FO} \over m_\DM } \times {s'\over s} \simeq \left({T'_\text{FO} \, \over m_\DM } \right) \left( { \eta^{3} g'_* \over g_*}\right)^{1\over 4} \left({ T_\text{RH} \over m_\text{Pl}}\right)^{9\over 4} \notag \\
&\simeq & 0.13 \left({ T'_\text{FO} \, \over m_\DM } \right) \left({ T_\text{RH} \over m_\text{Pl}}\right)^{9\over 4} \,.\label{estimate:nonrelaFO}
\end{eqnarray}
Notice that there is an undetermined factor in comparison with the relativistic case: the ratio of freeze-out temperature of the DM sector $T'_\text{FO}$ to DM mass. 
 
\end{itemize}

Determining which regime takes place requires one to know the freeze-out temperature, which can be calculated from the cross section of the 3-to-2 process~\cite{Bernal:2015ova}. Nevertheless, the numerical result for $Y_f$ does not significantly differ between both cases because the dependence on the reheating temperature is the same and the ratio $m_\DM/T'_\text {FO}$ generally gets a value of the order of 1--10 for the parameter region of our interest.

Now by comparing Eqs.~(\ref{estimate:relaFO} and \ref{estimate:nonrelaFO}) above with Eq.~\eqref{estimate:nonthermal} for the freeze-in mechanism, one can see that for $m_\DM \ll T_\text{RH} \ll m_\text{Pl}$, introducing dark thermalization leads to an enhanced production of DM particles by a factor around $(m_\text{Pl} / T_\text{RH})^{3/4}$. 
As is known, the reheating temperature is argued to be bounded from above by $3.2\times 10^{16}\,r_\star^{1/4}$\,GeV~\cite{Weinberg:2008zzc}, where $r_\star$ is the tensor-to-scalar ratio in the primordial fluctuations. 
The latter is in turn constrained to be smaller than 0.12 at 95\% C.L. by BICEP2/Keck and Planck data~\cite{Ade:2015tva}. Therefore, in contrast to freeze-in, the dark freeze-out scenario allows for DM masses as low as tens of MeV while satisfying the observed value of DM abundance. 

All our results are summarized in Fig.~\ref{fig:plot1}. There, the color gradient shows the required reheating temperature under the assumption of $m_\DM/T'_\text {FO}\sim 10$ and that the DM relic density matches the observed value. 
Note that $m_\DM/T'_\text {FO}$ in general varies within a very narrow range, only mildly modifying the result of Fig.~\ref{fig:plot1}. The left gray region, corresponding to $T_\text{RH} \gtrsim 10^{16}$\,GeV, is considered to be strongly disfavored.

Finally, we would like to remark that, for the parameter space of Fig.~\ref{fig:plot1} and according to Eq.~\eqref{eq:ind}, the lifetime of the spin-2 DM particle is at least $\unit[10^{50}]{s}$, which is well beyond the detection sensitivity of current indirect searches of DM.

\section{Conclusions}
\label{sec:conclusion}

In this article, using bimetric gravity as a general framework, we have studied the self-interaction of spin-2 DM particles, and its possible role in both producing the observed DM abundance and addressing small-scale problems. Non-negligible self-interactions arise from requiring suppression of possible modifications of the well-tested GR predictions. 
The same requirement simultaneously guarantees the stability of the DM candidate, in agreement with current experimental data from indirect searches.

Quantitatively, the parameter region of interest corresponds to DM masses at MeV--GeV scale, and to values of the Planck-mass ratio, $\alpha$, well below $10^{-15}$ (see Eq.~\eqref{eq:action}). This region was previously deemed to be impossible for successful DM production. This is because, on the one hand, the corresponding parameter $\alpha$ --which also plays the role of coupling constant between the DM and SM fields-- is too weak to achieve thermal freeze-out. On the other hand, because such DM masses are too small to generate the correct abundance from freeze-in without violating known experimental constraints, especially the upper bounds on reheating temperature, gravitational-waves energy density and large-scale isocurvature perturbations, as derived from high-precision cosmological data. 

The essence of this work is to point out that the presence of DM self-interactions for the parameter region of our interest leads to the self-thermalization of the dark sector, which in turn opens up the possibility of another production mechanism for spin-2 DM: the dark freeze-out. In this mechanism, the number-changing process 2DM\,$\leftrightarrow$\,3DM can significantly enhance the number of DM particles at an earlier time with respect to freeze-in of the same total energy, and allow for entropy-conserving freeze-out within the dark sector at a later time. 

We have then solved the dark freeze-out evolution of such spin-2 DM candidate. Concretely, for each DM mass from MeV to GeV scale, in Fig.~\ref{fig:plot1} we have shown the corresponding values of the parameter $\alpha$ as well as of the reheating temperature that produce the observed relic abundance. Interestingly, self-scattering cross sections per mass, $\sigma_\text{SI} /m_\text{FP}$, as large as $\mathcal{O}(1)$\,cm$^2$/g are still experimentally allowed. It means that the light DM candidate originated from bimetric gravity is also capable of addressing the long-standing astrophysical puzzles observed in the inner region of dark halos. 

At last, we would like to point out that this model of spin-2 DM is very sensitive to the reheating temperature. On the one hand, a determination of the reheating temperature by future experimental observations, e.g., by detecting the ratio of tensor-to-scalar primordial fluctuations, will also determine the DM mass to a large degree. On the other hand, such a light DM would be strongly disfavored if the upper bound on $T_\text{RH}$ is improved by one or two orders of magnitude.

\section*{Acknowledgments}

The authors would like to thank C\'eline Boehm, Subir Sarkar, Angnis Schmidt-May and Federico Urban for useful discussions. X.C. is supported by the ‘New Frontiers’ program of the Austrian Academy of Sciences. C.G.C. is supported by the Belgian Federal Science Policy through the Interuniversity Attraction Pole P7/37.

\appendix

\begin{widetext}
\section{Cross section formulae}
\label{sec:app}
For calculating the self-interaction cross section, we first determine the quadratic, cubic and quartic terms of the action in Eq.~\eqref{eq:action} where $\delta M$ is involved. At the leading order in $\alpha$, such terms turn out to be independent of $\delta G$. For the Einstein-Hilbert action of the metric $f$, we obtain
\begin{eqnarray}
{\cal S}^\text{EH} = m_g^2 \alpha^2 \int d^4 x \sqrt{-f} R(f) 
 \to\int d^4 x \left( {\cal L}_2^\text{EH}+{\cal L}_3^\text{EH}+{\cal L}_4^\text{EH}+\ldots \right)\,,
\end{eqnarray}
with
\begin{eqnarray}
{\cal L}_2^\text{EH} &=& \left.-\frac{1}{2}
 \,\partial_{\alpha }\delta M^{\alpha }{}_{\beta } \,\partial^{\beta }\text{tr}(\delta M)+\frac{1}{2} \,\partial_{\alpha }\delta M_{\beta \mu } \,\partial^{\beta }\delta M^{\alpha \mu }
-\frac{1}{4}\,\partial_{\alpha }\delta M_{\beta \mu } \,\partial^{\alpha }\delta M^{\beta \mu
 }+\frac{1}{4}\,\partial^{\alpha }\text{tr}(\delta M) \,\partial_{\alpha }\text{tr}(\delta M)\right.
\label{eq:kinHE}\,,
\\
{\cal L}_3^\text{EH}&=& 
\frac{1}{4 \,\alpha\, \mPl} \delta M_{\alpha \beta } \left(
-2 \,\partial_{\mu }\delta M^{\alpha }{}_{\nu } \,\partial^{\nu }\delta M^{\beta \mu }
+2 \,\partial_{\mu }\delta M^{\alpha }{}_{\nu } \,\partial^{\mu }\delta M^{\beta \nu }-2 \,\partial^{\mu }\text{tr}(\delta M) \,\partial_{\mu
 }\delta M^{\alpha \beta }
\right.\,\nonumber\\&& \left.
+2 \,\partial^{\alpha }\text{tr}(\delta M) \,\partial_{\nu }\delta M^{\beta \nu }
+2 \,\partial^{\mu }\text{tr}(\delta M) \,\partial^{\alpha }\delta M^{\beta }{}_{\mu }
-4 \,\partial^{\alpha }\delta M_{\mu \nu } \,\partial^{\mu }\delta M^{\beta \nu}
+\,\partial^{\alpha }\delta M_{\mu \nu } \,\partial^{\beta }\delta M^{\mu \nu }
\right.\,\nonumber\\&& \left.
-\,\partial^{\alpha }\text{tr}(\delta M) \,\partial^{\beta }\text{tr}(\delta M)
+2 \,\partial_{\nu }\delta M^{\mu \nu } \,\partial_{\mu }\delta M^{\alpha \beta }\right) 
+\frac{1}{2\,\alpha\,\mPl} \text{tr}(\delta M) {\cal L}^\text{EH}_2 \,,
\label{eq:3HE}
\\
{\cal L}_4^\text{EH}&=&
-\frac{1}{4\alpha^2 \mPl^2} \delta M_{\alpha \beta } \left(-2 \delta M^{\alpha }{}_{\mu } \,\partial_{\nu }\delta M^{\beta }{}_{\rho } \,\partial^{\rho }\delta M^{\mu \nu }
-\delta M_{\mu \nu } \,\partial_{\rho }\delta M^{\alpha \beta } \,\partial^{\rho }\delta M^{\mu \nu}
+\delta M_{\mu \nu } \,\partial_{\rho }\delta M^{\alpha \mu } \,\partial^{\rho }\delta M^{\beta \nu }
\right.\,\nonumber\\&& \left.
-2 \delta M^{\alpha }{}_{\mu } \,\partial^{\nu }\text{tr}(\delta M) \,\partial_{\nu }\delta M^{\beta \mu }
+2 \delta M^{\alpha }{}_{\mu } \,\partial_{\nu }\delta M^{\beta }{}_{\rho
 } \,\partial^{\nu }\delta M^{\mu \rho }
+2 \delta M^{\alpha }{}_{\mu } \,\partial^{\beta }\text{tr}(\delta M) \,\partial_{\rho }\delta M^{\mu \rho }
\right.\,\nonumber\\&& \left.
+2 \delta M_{\mu \nu } \,\partial_{\rho }\delta M^{\beta \rho } \,\partial^{\alpha }\delta M^{\mu \nu }
-4 \delta M_{\mu \nu }
 \,\partial^{\alpha }\delta M^{\mu }{}_{\rho } \,\partial^{\rho }\delta M^{\beta \nu }
+2 \delta M_{\mu \nu } \,\partial^{\alpha }\delta M^{\beta }{}_{\rho } \,\partial^{\rho }\delta M^{\mu \nu }
\right.\,\nonumber\\&& \left.
-4 \delta M^{\alpha }{}_{\mu } \,\partial^{\beta }\delta M_{\nu \rho } \,\partial^{\nu
 }\delta M^{\mu \rho }
+2 \delta M^{\alpha }{}_{\mu } \,\partial^{\nu }\text{tr}(\delta M) \,\partial^{\beta }\delta M^{\mu }{}_{\nu }
-2 \delta M_{\mu \nu } \,\partial^{\alpha }\delta M^{\mu }{}_{\rho } \,\partial^{\nu }\delta M^{\beta \rho }
\right.\,\nonumber\\&& \left.
+2 \delta M_{\mu \nu } \,\partial^{\nu
 }\text{tr}(\delta M) \,\partial^{\alpha }\delta M^{\beta \mu }
+\delta M^{\alpha }{}_{\mu } \,\partial^{\beta }\delta M_{\nu \rho } \,\partial^{\mu }\delta M^{\nu \rho }
-\delta M^{\alpha }{}_{\mu } \,\partial^{\beta }\text{tr}(\delta M) \,\partial^{\mu }\text{tr}(\delta M)
\right.\,\nonumber\\&& \left.
+2 \delta M_{\mu 
 \nu } \,\partial^{\alpha }\delta M^{\mu }{}_{\rho } \,\partial^{\beta }\delta M^{\nu \rho }
-2 \delta M_{\mu \nu } \,\partial^{\beta }\text{tr}(\delta M) \,\partial^{\alpha }\delta M^{\mu \nu }
+2 \delta M^{\alpha }{}_{\mu } \,\partial_{\rho }\delta M^{\nu \rho } \,\partial_{\nu
 }\delta M^{\beta \mu }\right) 
\,\nonumber\\&&
 -\frac{1}{8\alpha^2 \mPl^2} \left( \tr{\delta M}^2+2 \tr{\delta M^2} \right) {\cal L}_2^\text{EH}
+\frac{1}{2\alpha\mPl }\tr{\delta M}{\cal L}^\text{EH}_3\,.
\label{eq:4HE}
\end{eqnarray}

where the indices are raised and lowered with the background metric. Similarly, the potential in Eq.~\eqref{eq:action} gives
\begin{eqnarray}
{\cal S}^\text{pot} = -2m_g^4 \alpha^2 \int d^4 x \sqrt{-g} V\left(\sqrt{g^{-1}f}\right) 
 \to\int d^4 x \left( {\cal L}_2^\text{pot}+{\cal L}_3^\text{pot}+{\cal L}_4^\text{pot}+\ldots \right)\,,
\end{eqnarray}
with
\begin{eqnarray}
{\cal L}^\text{pot}_2 &=& -\frac{\mPl^2}{4} \left(\beta_1+2\beta_2+\beta_3\right)\left(\tr{\delta M^2}-\tr{\delta M}^2\right)\,, \label{eq:2pot} \\
{\cal L}^\text{pot}_3 &=& \frac{\mPl}{\alpha} \left(\left(\frac{5\, \beta_1+12 \,\beta_2+7 \,\beta_3}{24 } \right)\tr{\delta M^3}
-\left(\frac{2\, \beta_1+5 \,\beta_2+3\, \beta_3}{8}\right)\tr{\delta M}\tr{\delta M^2}
\right.\,\nonumber\\&&\left.
+\left(\frac{\beta _1+3 \beta _2+2 \beta _3}{24 }\right)\tr{\delta M}^3
\right)\,, \label{eq:3pot}\\
\label{eq:4pot}
{\cal L}^\text{pot}_4 &=&\frac{1}{\alpha^2}\left(
-\frac{1}{64} \left(11 \beta _1+28 \beta _2+19 \beta _3\right) \text{tr}\left(\delta M^4\right)
+\frac{1}{48} \left(8 \beta _1+21 \beta _2+15 \beta _3\right) \text{tr}\left(\delta M^3\right)
 \text{tr}(\delta M)
\right.\,\\&&\left.
-\frac{1}{32} \left(2 \beta _1+6 \beta _2+5 \beta _3\right) \text{tr}\left(\delta M^2\right) \text{tr}(\delta M)^2
+\frac{1}{64} \left(4 \beta _1+11 \beta _2+8 \beta _3\right)
 \text{tr}\left(\delta M^2\right)^2
\right.\,\nonumber\\&&\left.
+\frac{1}{192} \left(\beta _1+3 \beta _2+3 \beta _3\right) \text{tr}(\delta M)^4\right) .\nonumber
\end{eqnarray}
Notice that these expressions are independent of $\delta G$ as well. Finally, we would like to remark that the first term of Eq.~\eqref{eq:action} -that is, the Einstein-Hilbert action for the metric g- does not contribute the previous Lagrangians to leading order in $\alpha$. 

The quadratic and cubic pieces of the action that we obtain agree with those of Ref.~\cite{Babichev:2016bxi}. The quartic pieces, crucial for calculating the self-interaction rates, have not been reported in the literature to the best of our knowledge. Moreover, we also calculate the quintic terms and find they are proportional to $1/(\alpha^3 \mPl^{n+1})$, where $n=2$ for the EH piece --always containing two derivatives-- as well as $n=0$ for the potential piece which has no derivatives.

We have implemented Eqs.~(\ref{eq:kinHE}--\ref{eq:4HE}) and (\ref{eq:2pot}-\ref{eq:4pot}) in FeynRules~\cite{Alloul:2013bka} in order to derive the corresponding vertices. Then, we use them to calculate the nonrelativistic self-interaction cross section, and find

{\small
\begin{align}
\sigma_\text{SI} &= \frac{(\beta_1+2\beta_2+\beta_3)^2}{\pi \alpha^4 m_\DM^2} F\left(\frac{\beta_1-\beta_3}{\beta_1+2\beta_2+\beta_3}\right)\,,
&\text{with}\hspace{10pt}
&
F(r)= \frac{433 r^4-15756 r^3+214524
 r^2-1352304r+3610512}{53084160 }\, \,
\label{eq:Fr}
\end{align}
}

To double-check our results, we decompose the cross section into the contributions associated with different eigenvalues of the total angular momentum $J$. Introducing 
$r = \frac{\beta_1-\beta_3}{\beta_1+2\beta_2+\beta_3}$, we find 
\begin{eqnarray}
\sigma_\text{SI} &=& \sum^4_{J=0} (2J+1)\sigma_J\,,
\end{eqnarray}
where
\begin{align}
\sigma_0&=\frac{m_\DM^2 \left(13 r^2-206
 r+548\right)^2}{29491200 \pi \alpha ^4
 \mPl^4}\,,&
\sigma_2=\frac{m_\DM^2 \left(2 r^2-51
 r+342\right)^2}{16588800 \pi \alpha ^4
 \mPl^4}\,,&&
\sigma_4=\frac{m_\DM^2 \left(r^2-22
 r+136\right)^2}{7372800 \pi \alpha ^4
 \mPl^4}\,
\end{align}
and $\sigma_1= \sigma_3 =0$, as expected from spin statistics. 

\end{widetext}

\bibliographystyle{apsrev}
\bibliography{BIB}
\end{document}